\documentclass[aps, amsmath, preprint, amssymb,groupedaddress]{revtex4}

\setcitestyle{super}
\usepackage{graphicx}
\usepackage{gensymb}
\pdfoutput=1
\usepackage{amsmath}
\usepackage{color}
\usepackage{float}
\usepackage[english]{babel}
\usepackage{lipsum}
\usepackage{color,soul}
\begin{document}
\title{Dynamic Surface Modification due to Effusion of Na in Na$_2$IrO$_3$}
\author{Aastha Vasdev}
\thanks{These two authors contributed equally}
\author{Lalit Yadav}
\thanks{These two authors contributed equally}
\author{Suman Kamboj}
\author{Kavita Mehlawat}
\author{Yogesh Singh}
\author{Goutam Sheet}
\email{goutam@iisermohali.ac.in}
\affiliation{Department of Physical Sciences,  
Indian Institute of Science Education and Research Mohali (IISER M),
Sector 81, S. A. S. Nagar, Manauli, PO: 140306, India}
\begin{abstract}
The honeycomb lattice iridate Na$_2$IrO$_3$ shows frustrated magnetism and can potentially display Kitaev-like exchange interactions. Recently, it was shown that the electronic properties of the surface of crystalline Na$_2$IrO$_3$ can be tuned by Ar plasma treatment in a controlled manner leading to various phases of matter ranging from a fully gapped to a metallic surface, where the possibility of a charge-density wave (CDW) like transition is also expected. Here, through direct imaging with an atomic force microscope (AFM) in air, we show that the surface of crystalline Na$_2$IrO$_3$ evolves rapidly as elemental Na effuses out of the interleave planes to the surface and undergoes sublimation thereby disappearing from the surface gradually over time. Using conductive AFM we recorded a series of topographs and surface current maps simultaneously and found that the modification of the surface leads to change in the electronic properties in a dynamic fashion until the whole system reaches a dynamic equilibrium. These observations are important in the context of the exotic electronic and magnetic properties that the surface of Na$_2$IrO$_3$ displays.

\end{abstract}
\maketitle
Na$_2$IrO$_3$ has been recognized as a promising playground for studying various strongly correlated phenomena\cite{yogesh1,cao,choi}. Owing to its potential for showing novel ground states driven by the interplay of spin orbit coupling and strong electron correlations\cite{G.Jackeli,J.Chaloupka,wan,krempa}, it has drawn considerable attention of the condensed matter community and inspired a great deal of theoretical and experimental efforts\cite{pesin,yogesh1,liu,YogeshSingh4,reuther,yu}. Na$_2$IrO$_3$ has a layered honeycomb structure where oxygen mediated super-exchange between the Ir$^{4+}$  moments is expected to result in practical realization of a Kitaev model system \cite{cao,J.Chaloupka}. Depending on the relative strength of the spin-orbit coupling term in the Hamiltonian, it can show different ground states including conventional Neel order and quantum spin liquid behaviour with excitations comprising of the hitherto elusive Majorana fermions \cite{cao,ChoongH.Kim,YogeshSingh4,choi,A.Kitaev}. Under certain conditions, Na$_2$IrO$_3$ has also been proposed to be a topological insulator, a material with a bulk band gap and topologically protected conducting surface states thereby becoming a possible candidate to show quantum spin Hall (QSH) effect at relatively higher temperatures\cite{kimchi,Shitade}. As expected, the physical properties of Na$_2$IrO$_3$ depend on variation of the crystal structure, especially the change in interlayer distance and the nearest neighbour interaction strength\cite{fengye}. First principle calculations based on effective tight binding model have shown that a small change in the interlayer distance of Na$_2$IrO$_3$ or interaction strength can drive a regular band insulator phase of Na$_2$IrO$_3$ to a topological insulating phase through a quantum phase transition\cite{Yamaji}. On experimental side, recently it has been reported that the electronic properties of the surface of Na$_2$IrO$_3$ single crystals can be tuned by doping through Ar plasma treatment in a controlled manner ranging from a fully gapped to a metallic surface, where the possibility of a charge-density wave (CDW) like instability is also speculated\cite{Yogesh SinghAr}.  Hence, it is of significant importance to study the surface properties\cite{Cava} of Na$_2$IrO$_3$ crystals in detail. In this paper, we report detailed investigation of the physical, chemical and electronic properties of the surface of high quality Na$_2$IrO$_3$ crystals using scanning probe microscopy (SPM) in various modes.

High quality single crystals were used for the measurements presented in this paper. The crystals were available in the form of flakes with layers that can be easily cleaved mechanically. The freshly cleaved samples were mounted on the sample stage of an atomic force microscope (MFP 3D of Asylum Research) within 2-3 minutes of cleaving and the cleaved surfaces were imaged in non-contact mode using Si cantilevers with Pt-Ir coating. The spring constant of the cantilevers varied in a small range around 1.0 N/m. An AFM topograph of a freshly cleaved sample surface is shown in Figure 1(a). The tiny granular structures are blobs of metallic sodium (Na) that have diffused out from the layers underneath. The average size of the grains is approximately 8 nm and they continuously grow in size. As shown in Figure 1(b), the size of the blobs have almost doubled after 10 hours and the average size of the grains are seen to be approximately 15 nm. Beyond 10 hours, some of the blobs started coalescing with each other thereby giving rise to a distribution of particle size over the same scan area. Several blobs with average height larger than 20 nm can be seen to have grown after 20 hours of continuous scanning (Figure 1(c)). The size of the grains keep growing and as shown in Figure 1(d) the average size of the blobs have become more than 30 nm after 30 hours and the coalescing effect is more prominently visible in images captured after 40 hours (figure 1(e)) and 50 hours (figure 1(f)) of cleaving. Imaging of the same area was continuously done for 70 hours and the gradual increase in blob size can be clearly seen in the movie file provided as supplemental material. It can be seen that the blob size does not change noticeably beyond 50 hours. 

\begin{figure}[h!]
\begin{center}
\includegraphics[scale=1]{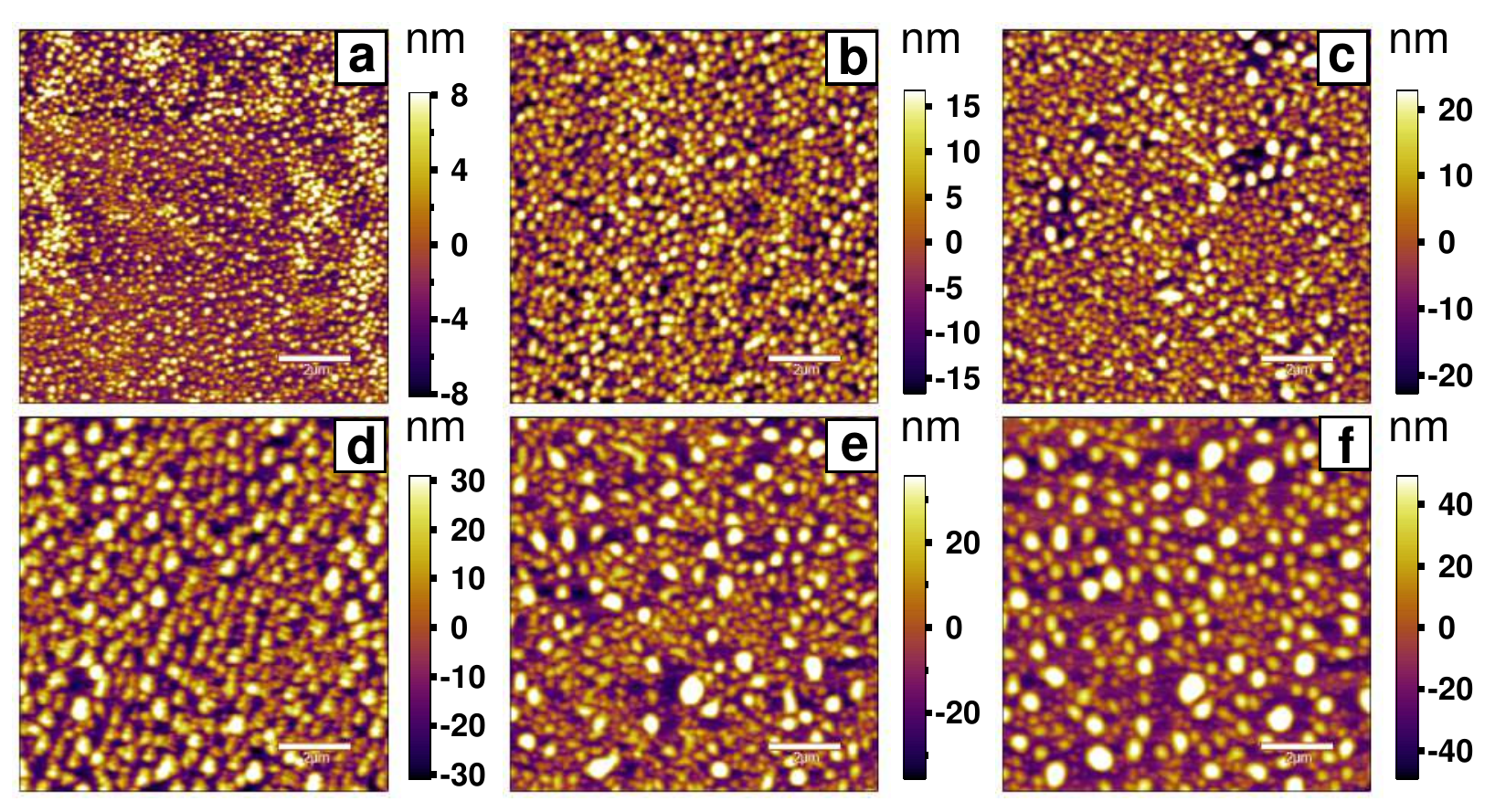}
\caption[]{Evolution of surface topography of a single crystal of Na$_2$IrO$_3$ with time. Topographs show the emergence of sodium blobs scanned at an interval of (a)0 hrs (b)10  hrs (c)20 hrs (d)30 hrs (e)40 hrs (f)50 hrs after cleaning}
\label{Serie}
\end{center}
\end{figure}

EDX analysis of the surface revealed that while in the clean area of the surface, ratio of Na and Ir was 2:1 as expected for Na$_2$IrO$_3$, the amount of Ir on the blobs is negligible compared to that of Na. Since EDX is a surface sensitive characterizing technique, it can be inferred that the blobs are primarily composed of sodium and the absence of Iridium points towards the fact that elemental sodium has effused out from the layers underneath.

The effusing Na clusters cover the entire surface sufficiently long time after cleaving and it is natural to expect that they will contribute to surface conductivity of the crystal. In order to check whether the blobs on the top surface are electrically connected to the bottom surface, we performed conductivity mapping of the top surface of crystal. For such experiments, the bottom surface of the sample was grounded and a conducting tip, mounted at the end of a soft cantilever, was brought in contact with the top surface. A voltage bias of 10V was applied to the tip and local current variations along with topography were recorded. A current map corresponding to a topography of 8$\mu$m x 8$\mu$m area (Figure 2(a)) is shown in Figure 2(e).

\begin{figure}[h!]
\begin{center}
\includegraphics[scale=.9]{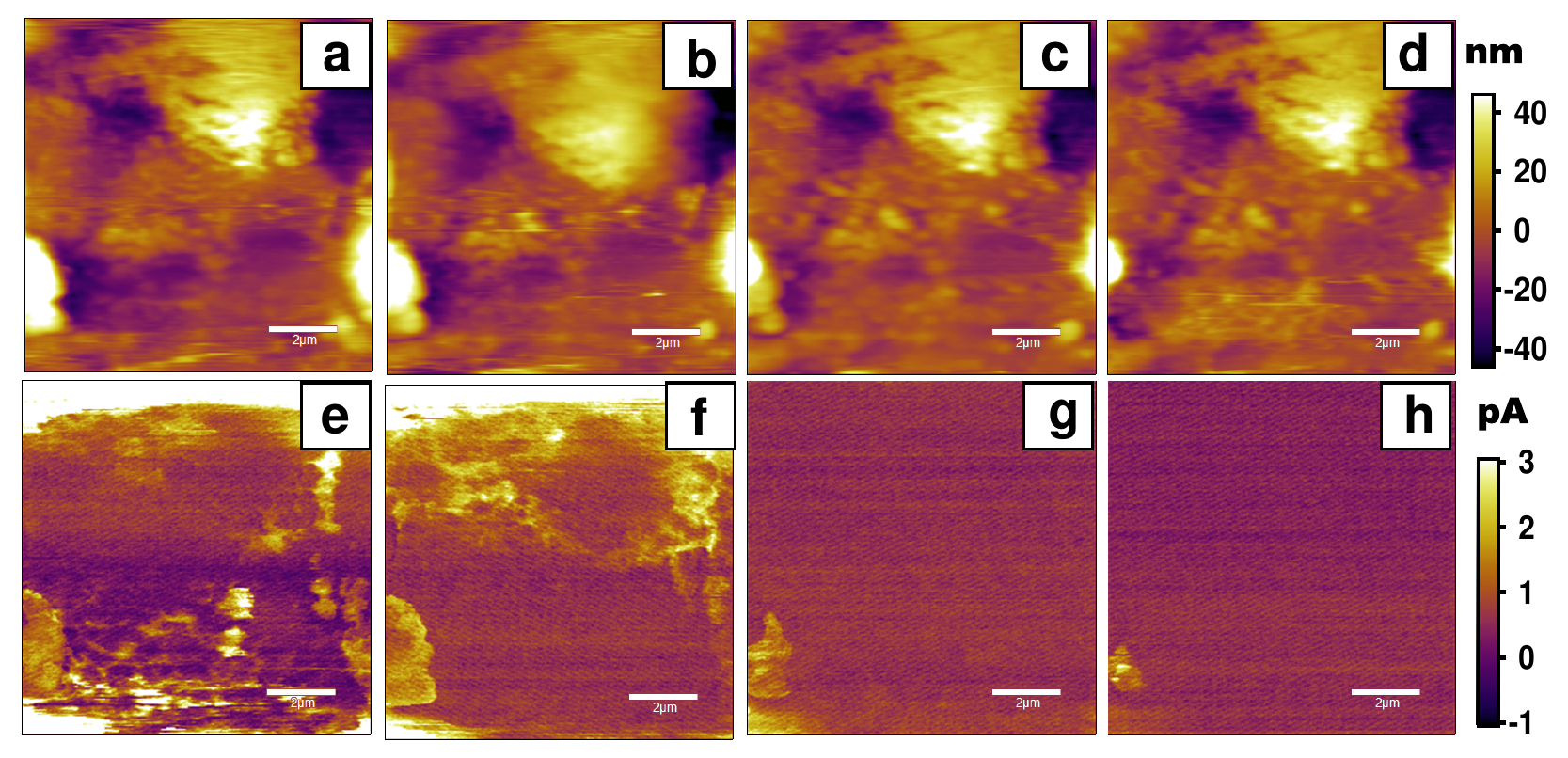}
\caption[Sèries temporals de vent i direccions]{Conductive AFM on an area of 8$\mu$m x 8$\mu$m by applying a bias of $10V$. (a), (b), (c) and (d) show the topography of the same area and (e), (f), (g) and (h) show corresponding current maps.}
\label{Serie}
\end{center}
\end{figure}
The bright regions in the current map correspond to the areas having more conductivity as compared to the dark regions. This means, sites having blobs of elemental sodium show higher conductivity in contrast to the background. Figure 2(b) and 2(c) are the topographs after first conductance map. Here it can be observed that as the tip scans the surface in contact mode, it modifies the surface by dragging the sodium to the edges of the scan area, thereby cleaning the area. Removal of sodium blobs  from the area is clearly seen in the current maps (Figure 2(f) and Figure 2(g)) where a sharp decrease in current is observed compared to that seen during first scan in contact mode. After four subsequent scans, most of the effused sodium was removed (Figure 2(d)) from the scan area and consequently current dropped to zero (Figure 2(h)).

	\paragraph*{•}
In order to further investigate the dynamics of the surface, we monitored the time evolution of the topographic features in the area scanned during the previous conductivity measurements (Figure 3(a)) from where Na blobs were removed by the tip. This was done continuously for next 40 hours. After 10 hours of scanning, we observed the sodium blobs emerging from the bottom left corner of the area under investigation (Figure 3(b)). The size of the blobs gradually grew in the next 10 hours (Figure 3(c)). 

\begin{figure}[h!]
\begin{center}
\includegraphics[scale=.5]{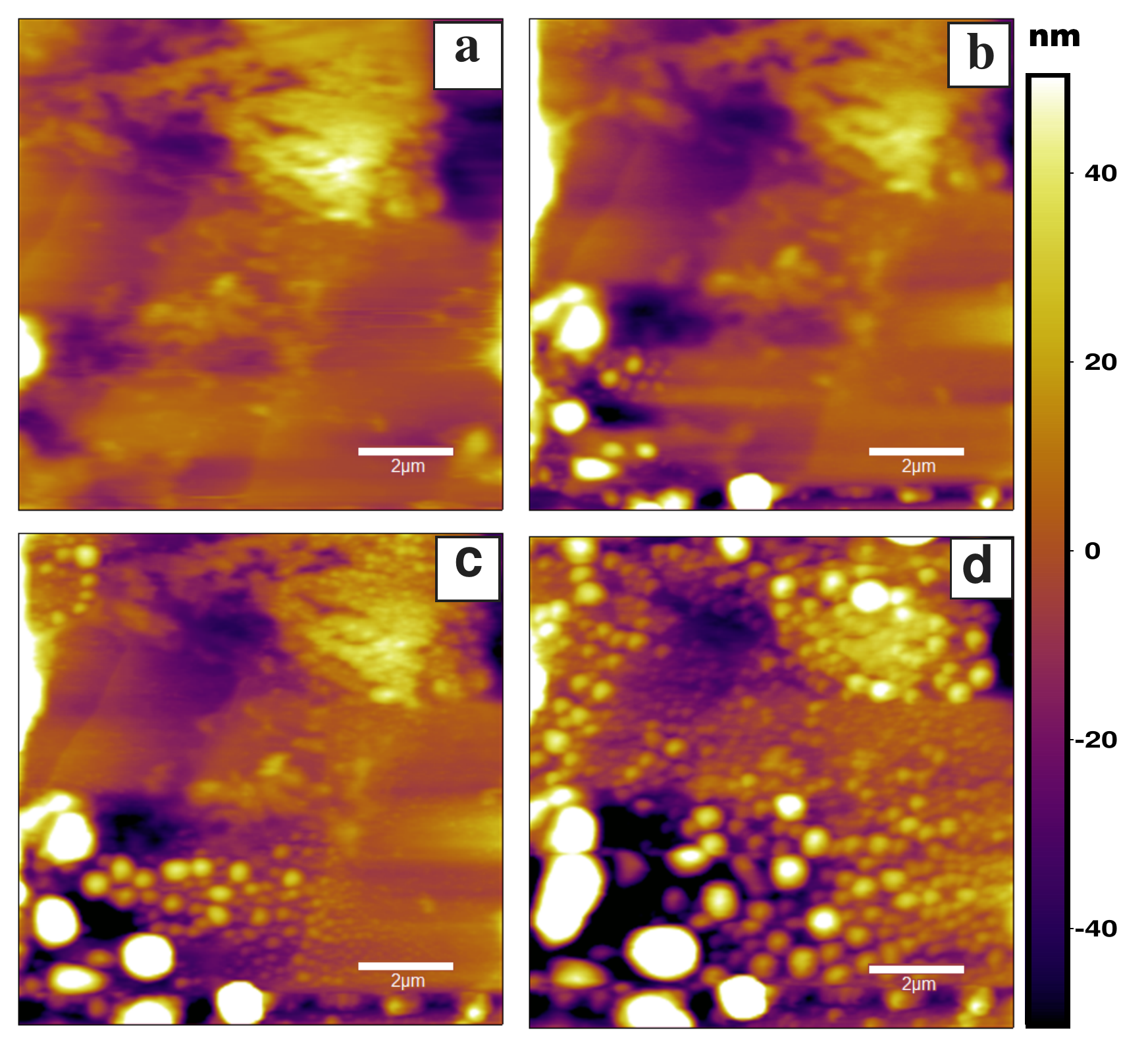}
\caption[Sèries temporals de vent i direccions]{Topography of the same scan area showing the emergence of sodium clusters. Images (a),  (b), (c) and (d) are observed at an interval of 10 hours each. }
\label{Serie}
\end{center}
\end{figure}
Fresh sodium blobs continued to emerge on the surface in other parts of the scan area and spread over the entire surface as it can be seen in the image captured after 40 hours (Figure 3(d)). This observation, as it is clear in the supplementary video-2, supports the idea that the elemental sodium continuously effuses from the underneath layers in Na$_2$IrO$_3$.

After around 50 hours, the growth of the blobs almost stopped possibly due to a dynamic equilibrium established between effusion and spontaneous sublimation of Na. After this situation was achieved, we performed conductive AFM measurements again by applying a voltage bias of 6V over the same area and obtained the current distribution as shown in Fig 4. In Fig 4(e), a current of 1-2pA is observed from the surface wherever sodium is present. As the Na is removed from the area by the tip due to multiple scans, the current dropped to zero again as shown in Fig 4(h).

\begin{figure}[h!]
\begin{center}
\includegraphics[scale=.9]{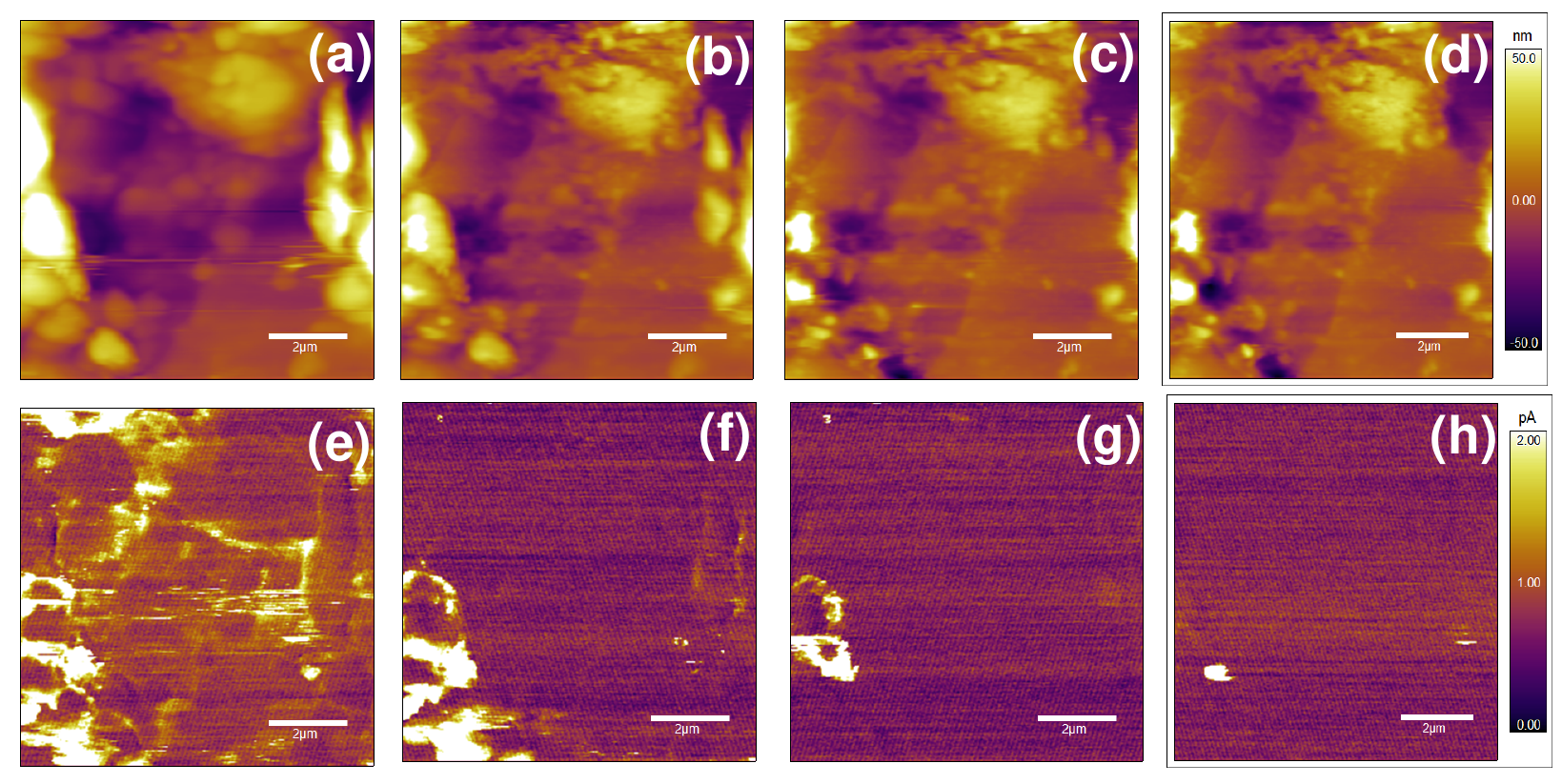}
\caption[Sèries temporals de vent i direccions]{Conductive AFM over the same area by applying a bias of 6V. (a), (b), (c) and (d) are the topographs while (e), (f), (g), (h) are the corresponding current maps of four successive scans.}
\label{Serie}
\end{center}
\end{figure}

\paragraph*{•}

The Na clusters formed on the area that was previously cleaned by the tip start disappearing from the surface after some time due to continued sublimation and exhaustion of effusive Na and as a consequence the blob size start reducing (Figure 5). The images (a), (b), (c), (d) in Fig 5 are the snapshots after  15 hours interval. The brighter grains visible in Fig 5(a) turn faint in the Fig 5(b) and (c) and finally disappear into background (Fig 5(d)).  
\begin{figure}[h!]
\begin{center}
\includegraphics[scale=.9]{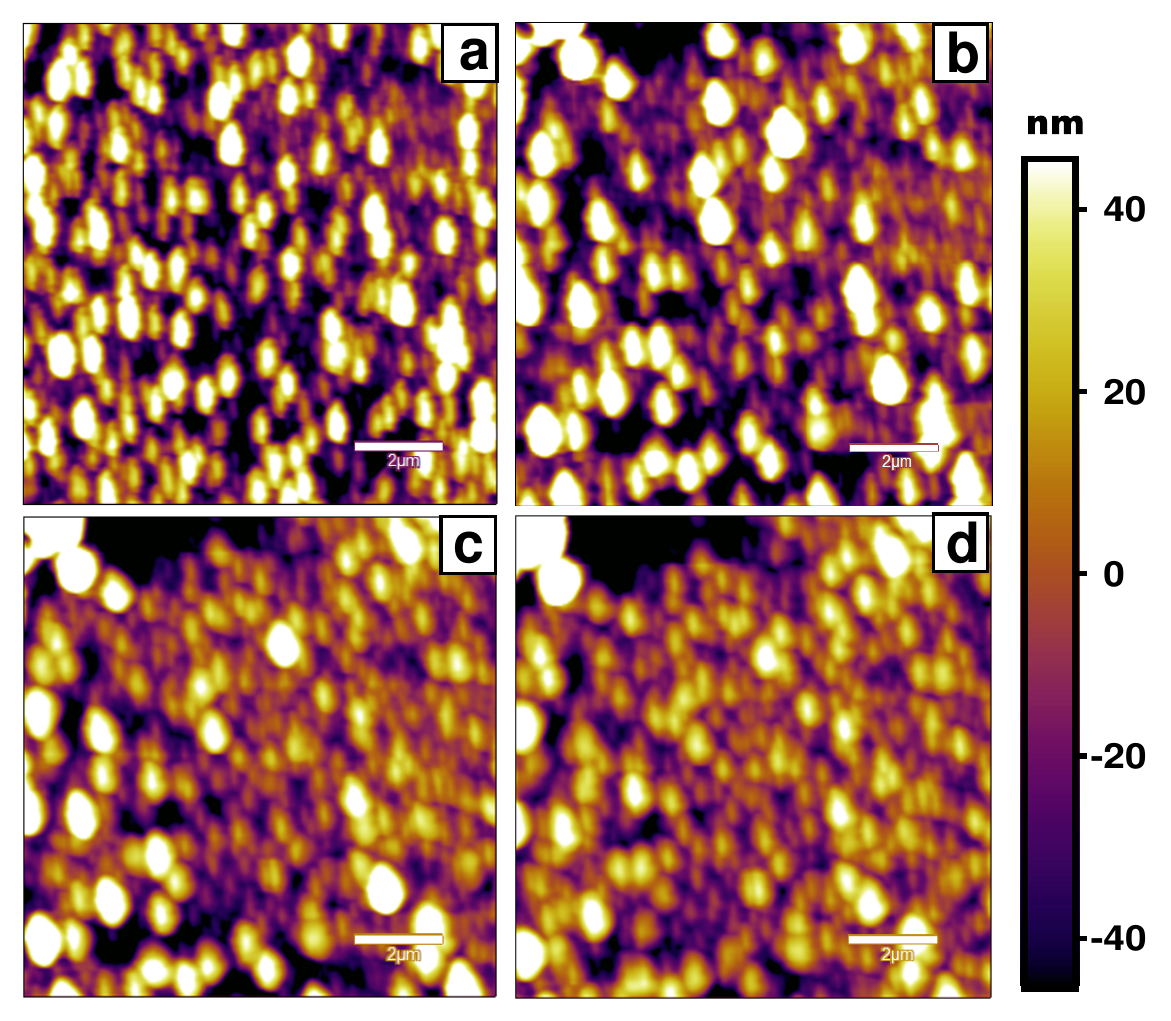}
\caption[Sèries temporals de vent i direccions]{Topography images showing decrease in the size of sodium blobs due to sublimation. (a), (b), (c) and (d) are the snapshots recorded at 15 hours interval.}
\label{Serie}
\end{center}
\end{figure}

In conclusion, we have demonstrated that the surface of Na$_2$IrO$_3$ crystals become Na rich due to effusion of Na from the bulk. The effusing sodium get accumulated in the form of clusters on the crystal surface.  Therefore, the observation of the exotic properties that Na$_2$IrO$_3$ can potentially exhibit might be prohibited due to the effusion of Na to the surface. We have also shown that the Na clusters on the surface can be cleaned mechanically. From scanning probe microscopy in different modes we have shown that such clusters of Na make the surface significantly conductive. If the crystals are cooled down right after cleaning, the effusion process can be minimized and its intrinsic properties can be retained for investigation.

GS acknowledges financial support from the research grants of Swarnajayanti fellowship awarded by the Department of Science and Technology (DST), Govt. of India under the grant number DST/SJF/PSA-01/2015-16.

\end{document}